\documentclass[aps,prl,twocolumn,groupedaddress]{revtex4}

\usepackage{graphicx}
\usepackage{dcolumn}
\usepackage{bm}
\usepackage{amsmath}
\usepackage{amssymb}
\usepackage{mathrsfs}
\usepackage{float}
\usepackage{latexsym}
\usepackage{epsfig}
\usepackage{amsbsy}
\usepackage{array}
\usepackage{amssymb}
\usepackage{bm}
\usepackage{appendix} 
\usepackage{color}
\usepackage{tikz-feynman}
\usepackage{fancyhdr}
\usepackage{ulem}
\usepackage{cancel}
\usepackage{booktabs}
\usepackage{multirow}
\usepackage{pifont}
\usetikzlibrary{calc}
\usepackage{tikz}
\usepackage[colorlinks=true,linkcolor=red,citecolor=blue,urlcolor=blue,]{hyperref}
\usepackage[doipre={doi:~}]{uri}

\graphicspath{{figures/}}

\begin{document}
	\bibliographystyle{plain}

    \title{Floquet-Bloch Oscillations and Intraband Zener Tunneling in an Oblique Spacetime Crystal}
	
	\author{Qiang Gao}
	\affiliation{Department of Physics, University of Texas at Austin, Texas, USA}
	
	\author{Qian Niu}
	\affiliation{Department of Physics, University of Texas at Austin, Texas, USA}
	
	\date{\today}
	
	\begin{abstract}
    We study an oblique spacetime crystal realized by a monoatomic crystal in which a sound wave propagates, and analyze its quasienergy band structure starting from a tight-binding Bloch band for the static crystal. We investigate Floquet-Bloch oscillations under an external field, which show different characteristics for different band topologies. We also discover intraband Zener tunneling beyond the adiabatic limit, which effectively converts between different band topologies.  Our results indicate the possibility of a quantum acoustoelectric generator that converts energy between the sound wave and a DC electric field in quantized units.

	\end{abstract}

	\maketitle
	
	\textit{\textbf{Introduction}.---} 
	Periodically driven quantum systems have a long history of study in physics, and have emerged in recent years as a new playground for novel topological properties~\cite{rechtsman2013photonic,lindner2011floquet,yao2017topological} and quantum materials engineering~\cite{oka2019floquet}.   There are also discussions on the tantalizing possibility of the spontaneous formation of time crystals~\cite{wilczek2012quantum,else2016floquet,zhang2017observation,autti2018observation} and spacetime crystals~\cite{li2012space}, adding new excitements to this field.  However, there is still much to be explored, and there are challenges in understanding the electronic dynamics in such systems~\cite{genske2015floquet,messer2018floquet}.   
	
	According to a recently reported symmetry classification~\cite{Xu2018Spacetime}, spacetime crystals fall into rectangular and oblique categories, depending on whether the system has separate translational symmetries in space and time.  Here we present an example for the latter, a monoatomic crystal in which a single mode of sound wave propagates.  One can still make a Floquet-Bloch analysis, but quasienergies and momenta are now defined modulo an oblique Brillouin zone, and the usual concepts of Bloch oscillations and Zener tunneling for Bloch bands can be essentially modified. We find Floquet-Bloch oscillations unraveling unusual types of band topologies. We then discuss intraband Zener tunneling, which cannot occur for a rectangular spacetime crystal, and the adiabatic conditions for the validation of realizing one particular band topology.  Our results indicate a novel mechanism for a quantum acoustoelectric generator that converts energy between the sound wave and a dc electric field.

	\textit{\textbf{Floquet-Bloch analysis for an oblique spacetime crystal}.---}  The oblique spacetime crystal considered here is a monoatomic crystal with sound waves propagating through it:
	\begin{equation}\label{OrigianlHamltonian}
		H(\boldsymbol{x},t)=\frac{-\hbar^2}{2M}\nabla_{\boldsymbol{x}}^{2}+ \sum_{\boldsymbol{R}} V\left(\boldsymbol{x}-\tilde{\boldsymbol{R}} \right)
	\end{equation} 
	with the atomic position being time-dependent
	$
	\tilde{\boldsymbol{R}}=\boldsymbol{R}-\boldsymbol{A} \cos(\boldsymbol{\kappa}\cdot\boldsymbol{R} -\Omega t).
	$
	Here the $(\boldsymbol{\kappa},\Omega)$ is the momentum and frequency of that sound wave, $|\boldsymbol{A}|$ is the oscillation amplitude and $\boldsymbol{R} = n_1\boldsymbol{a}_1+n_2\boldsymbol{a}_2+n_3\boldsymbol{a}_3$ labels the lattice sites. This Hamiltonian has the following translational symmetries: $H(\boldsymbol{x},t+2\pi/\Omega) = H(\boldsymbol{x},t) = H(\boldsymbol{x}+\boldsymbol{R},t+\boldsymbol{\kappa}\cdot\boldsymbol{R}/\Omega)$, which defines an oblique spacetime lattice with non-orthogonal lattice vectors: $(\boldsymbol{0},\frac{2\pi}{\Omega})$ and $(\boldsymbol{R},\frac{\boldsymbol{\kappa}\cdot\boldsymbol{R}}{\Omega})$. Those vectors determine the reciprocal lattice structure to be also oblique, characterized by vectors: ($\boldsymbol{\kappa},\Omega$) and ($\boldsymbol{G},0$), where $\boldsymbol{G}$ is the reciprocal lattice vector of the corresponding static crystal~\cite{footnote1}. (When the sound wave vector $\boldsymbol{\kappa}$ is rationally related to $\boldsymbol{G}$, one may adopt a superlattice point of view with a folded Brillouin zone, so that the system may be taken as a rectangular spacetime superlattice, but there will be seemingly `mysterious' band crossings due to the band folding.)
	
	The Floquet-Bloch band theory of the oblique spacetime crystal goes quite parallel to that for a rectangular spacetime crystal~\cite{gomez2013floquet}.
	The eigenstates that respect periodicities of the Hamiltonian satisfy the time-dependent Schrodinger equation:
	\begin{equation}\label{Schrodinger_eq}
		H(\boldsymbol{x},t)|\Psi(\boldsymbol{x},t)\rangle=i\hbar\partial_{t}|\Psi(\boldsymbol{x},t)\rangle.
	\end{equation}
	We consider that $|\boldsymbol{A}| \ll\text{lattice constants}$ and the effect of lattice vibration to leading orders in the amplitude $\boldsymbol{A}$:
	$
		H(\boldsymbol{x},t) = H_0(\boldsymbol{x}) + \boldsymbol{A}\cdot\boldsymbol{H}_1(\boldsymbol{x},t) + \cdots,
	$
	where $H_0$ is the Hamiltonian of the corresponding static crystal. To simplify matters, we assume that the electrons are all in the lowest Bloch band of the static crystal, which is well-separated from all other bands energetically so that mixing with them can be ignored when the lattice vibration is turned on. 
	
	Under these conditions, lattice vibrations can still mix a Bloch state \{$|\psi_{\boldsymbol{k}}(\boldsymbol{x})\rangle $\} of energy $\omega_g(\boldsymbol{k})$ in the lowest band with its phononic ``sidebands'', which differ with each other by an integer multiple of the phonon energy and momentum $(\Omega, \boldsymbol{\kappa})$. 
	In other words, we can choose the basis states to be the phonon replica of a Bloch state:
	\begin{equation}
	    \{|\Phi_{n,\boldsymbol{k}}(\boldsymbol{x},t)\rangle\equiv e^{-in\Omega t}|\psi_{\boldsymbol{k}+n\boldsymbol{\kappa}}(\boldsymbol{x})\rangle\},
	\end{equation}
	where $n$ is the replica index.  This basis can be made orthonormal under a new inner product defined as
	\begin{equation}
		\langle\langle \phi(\boldsymbol{x},t)|\psi(\boldsymbol{x},t) \rangle\rangle \equiv \frac{1}{T}\int_{0}^{T}dt\int d\boldsymbol{x}\phi^*(\boldsymbol{x},t)\psi(\boldsymbol{x},t)  
	\end{equation}
	where $T = 2\pi/\Omega$ is the Floquet time period. 
	
	Having the phonon replica basis, we can then expand the eigenstate in Eq.\eqref{Schrodinger_eq} as: 
	\begin{equation}\label{solution}
		|\Phi_{\omega,\boldsymbol{k}}(\boldsymbol{x},t)\rangle =\sum_{n}e^{-i\omega t}f_{\boldsymbol{k}}^{n}|\Phi_{n,\boldsymbol{k}}(\boldsymbol{x},t)\rangle,
	\end{equation}
	where $(\omega,\boldsymbol{k})$ are the quasienergy and quasimomentum, respectively. Since those two quantities are conserved modulo a Brillouin zone due to the periodicity in the reciprocal space, we can use them as characterizations for the eigenstates.
	Utilizing the orthonormal conditions of the basis functions, we then find that the coefficients $f_{\boldsymbol{k}}^{n}$ satisfy the following matrix eigenvalue equation:
	$
	\sum_n\mathcal{H}_{m,n}(\boldsymbol{k})f^n_{\boldsymbol{k}}= \hbar\omega f^m_{\boldsymbol{k}}
	$
	where the elements of the Kernel matrix $\mathcal{H}(\boldsymbol{k})$ are given by 
	\begin{equation}\label{kernel}
	    \mathcal{H}_{m,n}\equiv\langle\langle\Phi_{m,\boldsymbol{k}}(\boldsymbol{x},t)|H(\boldsymbol{x},t)|\Phi_{n,\boldsymbol{k}}(\boldsymbol{x},t)\rangle\rangle-\hbar n\Omega\delta_{mn}.
	\end{equation}
	In the static limit of $\boldsymbol{A}=\boldsymbol{0}$, the matrix $\mathcal{H}$ is diagonal with eigenvalues $\omega_n(\boldsymbol{k}) = \omega_g(\boldsymbol{k}+n\boldsymbol{\kappa})-n\Omega$, meaning that the quasienergy bands are just the original Bloch band $\omega_g(\boldsymbol{k})$ shifted by the reciprocal lattice vector ($\boldsymbol{\kappa},\Omega$).  When lattice vibration is turned on, off-diagonal elements of the Hamiltonian will appear, which can open gaps at places where the Bloch band crosses with its phonon replicas. From the calculation, we also find a general relation between different quasienergy bands:
	\begin{equation}
		\omega_{n+m}(\boldsymbol{k}) = \omega_m(\boldsymbol{k}+n\boldsymbol{\kappa})-n\Omega,
	\end{equation}
	which reflects the periodicity in the reciprocal space.
	
\begin{figure}[t]
		\centering
		\includegraphics[width=8cm]{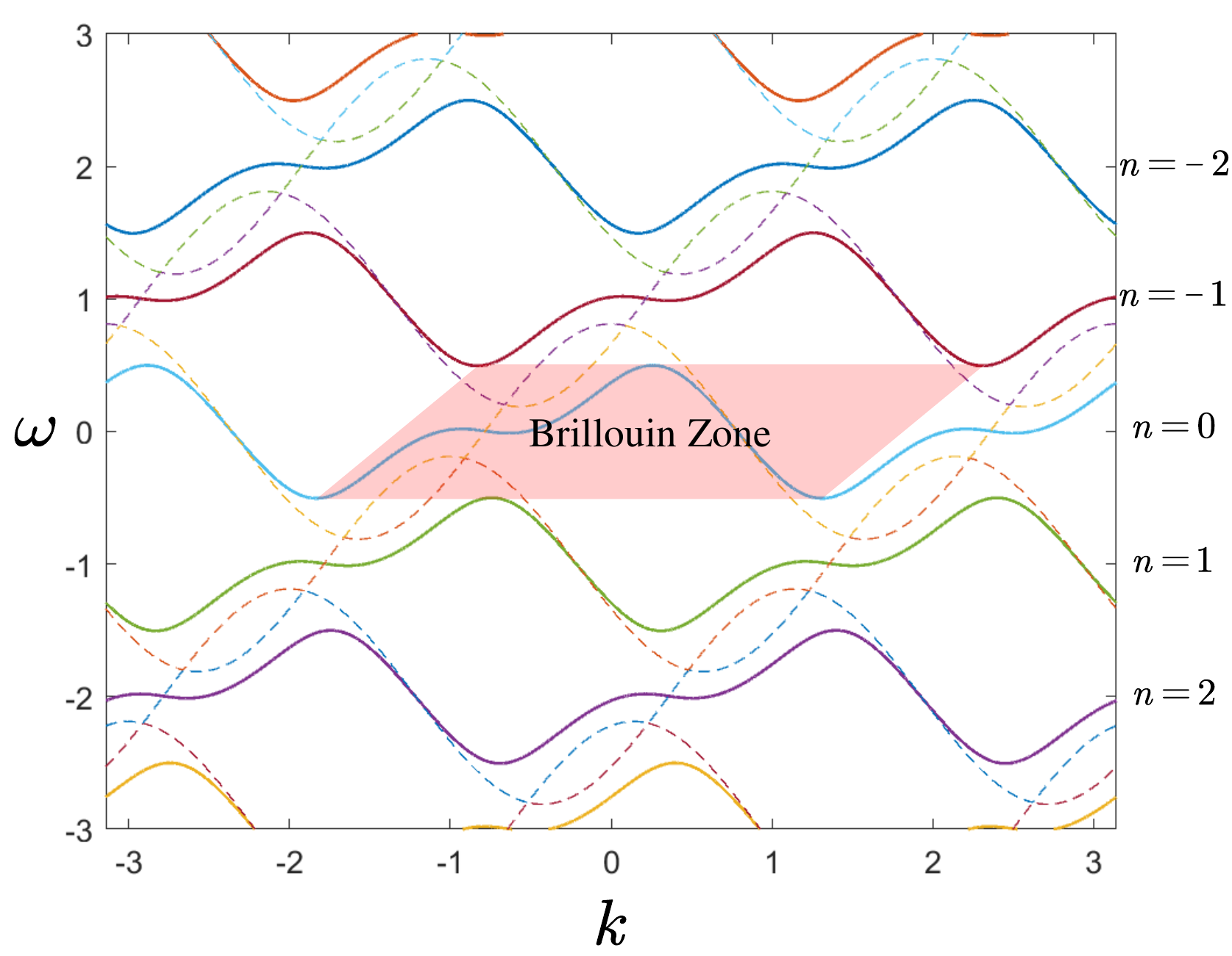}
		\caption{Floquet-Bloch band structure for the (1+1)D oblique spacetime crystal modeled by Eq.$\eqref{kernel}$, originating from the lowest Bloch band (dashed curves repeated over the Brillouin zone) of the unperturbed system. The quasienergy dispersion (solid curves) is calculated when a single mode of the sound wave is turned on to a finite amplitude~\cite{footnote2}.  The labels on the right vertical axis are the indices of the replicas of the same single band as defined within the Brillouin zone. }\label{band}
	\end{figure}

	These general features of the Floquet-Bloch band structure are illustrated in Fig.$\ref{band}$ for the case of a (1+1)D oblique crystal. The dashed curves are the unperturbed bands with no oscillation, and we can see they are nothing but replicas of the original cosine-shape Bloch band. The solid curves are the band dispersion under time-dependent perturbation. The red shaded area stands for the Brillouin zone of the oblique spacetime crystal characterized by two reciprocal lattice vectors: ($G=2\pi/a,0$) and ($\kappa,\Omega$). Without loss of physics, we take the region containing the ($n=0$) band $\omega_0(k)$ as our first Brillouin zone and all others as replicas.
	
	\textit{\textbf{Floquet-Bloch oscillations and band topology}.---} Very interesting phenomena such as Bloch oscillations and Zener tunneling can occur for Bloch bands in the presence of an external electric field $E$, and it is natural to ask what can happen to the quasienergy bands in a spacetime crystal. If we represent the field by a vector potential and treat its time dependence adiabatically, i.e., as very slow compared to all other time scales in the problem, we can still use the quasienergy states as solutions provided that the quasimomentum is replaced as $k\to k-eEt/\hbar$. Then by using the perturbation method, we can find an expression for the group velocity $\dot{x}$. Together with the time-dependent quasimomentum, the electronic motion in a quasienergy band can be summarized as 
	\begin{equation}\label{EOM}
		\begin{split}
			\dot{k} = -\frac{eE}{\hbar}, \quad
			\dot{x} = \frac{\partial \omega_n(k)}{\partial k},
		\end{split}
	\end{equation}
	which leads to a similar motion as Bloch oscillations that we call Floquet-Bloch oscillations in the present context~\cite{marder2010condensed}. Indeed, for the band structure in Fig.$\ref{band}$, the quasienergy is periodic in momentum, $\omega_n(k+2\pi/a) = \omega_n(k) $, which implies, according to the equations of motion, that the velocity of electrons is also periodic in time with period $\frac{\hbar}{eE}G$~\cite{dahan1996bloch}. Similar discussion in the time-driven system can be found in Ref.\cite{gagge2018bloch}.
	
	However, in the oblique spacetime crystal, quasienergy bands can also, in principle, exhibit nontrivial periodicity like $\omega_n(k+\kappa) = \omega_n(k)+\Omega$ or even more exotically $\omega_n(k+2\pi/a\pm\kappa) = \omega_n(k)\pm\Omega$, which will lead to new oscillation periods of $\frac{\hbar}{eE}\kappa$ and $\frac{\hbar}{eE}(G\pm\kappa)$, respectively. Those unique behaviors suggest different unusual band topologies~\cite{footnote3}.
	
\begin{figure}[t]
		\centering
		\includegraphics[width=8cm]{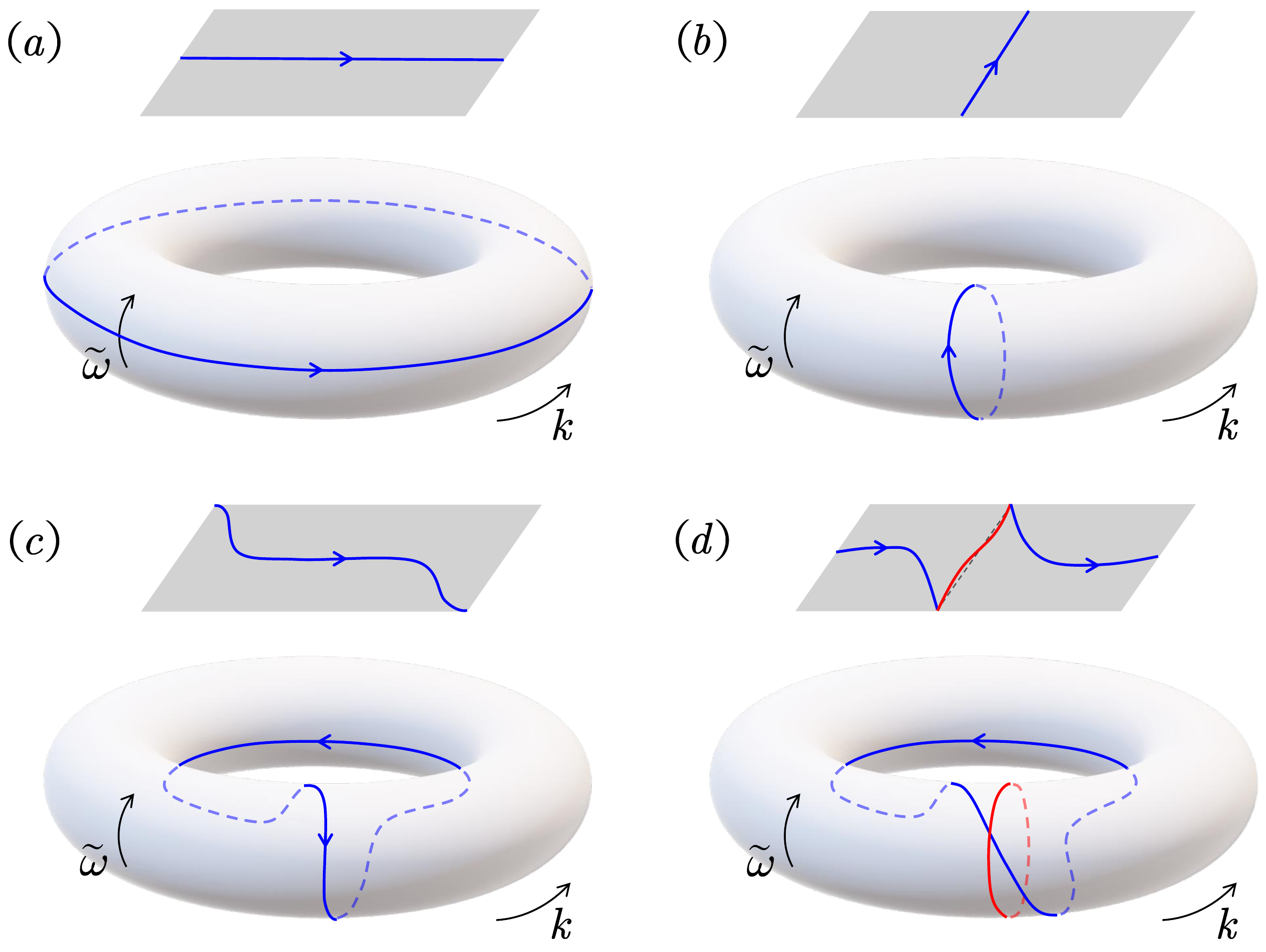}
		\caption{The topology of band dispersions (blue curves) as seen in the Brillouin zone and on the torus after shearing it and wrapping around along the reciprocal lattice vectors: $\boldsymbol{\tilde{\omega}}=(\kappa,\Omega)$ and $\boldsymbol{k}=(G,0)$. Panel (a) corresponds to the class shown in Fig.(1),  Panel (b) is a class that has yet to be seen and is only possible in oblique spacetime crystals~\cite{footnote4}.  Panel (c) corresponds to typical edge states of topological Floquet insulator in a higher dimension~\cite{yao2017topological}.  Panel (d) shows to a unique gapless band structure that combines the topologies in both (b,c).} \label{torus} 
	\end{figure}

    To appreciate the possibilities of different topologies,  we project the Brillouin zone onto a torus by shearing it into a rectangle and wrapping around to join the opposite edges.  A quasienergy dispersion is then characterized by a pair of winding numbers $N_{\omega}$ and $N_{k}$ around the two topologically distinct directions represented by the reciprocal lattice vectors $\boldsymbol{\tilde{\omega}}=(\kappa,\Omega)$ and $\boldsymbol{k}=(G,0)$.  The examples mentioned above are illustrated in Fig.$\ref{torus}$, with $N_{\omega} = 0, N_{k} = 1$ for panel (a), $N_{\omega} = 1, N_{k} = 0$ in panel (b), and $N_{\omega} = -1, N_{k} = 1$ in panel (c), corresponding to the Floquet-Bloch oscillations with periods $\propto N_kG+N_\omega \kappa$.  
    
	Although we do not yet find the possibilities illustrated in panels (b) and (c) within the model studied in this work, we discover a system called Oscillating Dirac Comb that can possess a gapless band structure for a specific oscillation amplitude (with more details given in the S.M.\cite{SM}), as shown schematically in panel (d). We have intentionally plotted the band in red and blue corresponding to the topologies in panels (b) and (c), respectively. Taking the band structure as two different topologies requires that the electron cannot be in a superposition of the two segments plotted in blue and red and must remain consistently on one of them when passing through the crossing point. 
	
	However, an exact gap closing in the oscillating Dirac comb requires a fine-tuning of parameters~\cite{SM}, which is not robust under any other perturbations and thus unrealistic in real experiments. So, we have to allow such system to have a tiny gap. In the next section, we will see how the joint topology shown in panel (d) is possible even with a tiny gap opened at the crossing point by discussing the intraband Zener tunneling and the adiabatic conditions.
	
	\textit{\textbf{Intraband Zener tunneling and adiabatic condition}.---} Zener tunneling refers to the breakdown of adiabaticity when the rate of parameter change cannot be regarded as small compared to the gap between the energy levels, and there is also an analog of the phenomenon between quasienergy levels in Floquet systems~\cite{breuer1989quantum,hijii2010symmetry,rodriguez2018floquet}.  In crystals under an electric field,
	the crystal momentum becomes a time-dependent parameter, and interband Zener tunneling has been well studied. Here, due to the fact that in oblique spacetime crystals, the gaps can be opened between a quasienergy band and its periodic replicas (as shown in Fig.~\ref{band}), we can actually anticipate an intraband Zener tunneling happening through such gaps.

	The analysis of the intraband Zener tunneling is quite similar to that of normal Zener tunneling between different Bloch bands. The key idea is that we consider tunneling between two eigenstates $|\psi_1(k)\rangle $ and $|\psi_2(k)\rangle $:
	\begin{equation}
		\begin{split}
			|\psi_1(k)\rangle =& \sum_n f_{k}^n |\Phi_{n,k}(x,t)\rangle \\
			|\psi_2(k)\rangle =& \sum_n f_{k-\kappa}^n |\Phi_{n-1,k}(x,t)\rangle,
		\end{split}
	\end{equation}
	sitting on two adjacent bands (replicas) labeled by 1 and 2, which have quasienergies $\epsilon_1=\omega(k)$ and $\epsilon_2=\omega(k-\kappa) + \Omega$, respectively, with a direct gap $\Delta_0$. One can check that $|\psi_1(k)\rangle$ and $|\psi_2(k)\rangle$ are orthonormal ($\langle\langle \psi_i(k)|\psi_j(k)\rangle\rangle = \delta_{ij} $). The reason why such tunneling is indeed an intraband process is that $|\psi_2(k)\rangle$ is equivalent to $|\psi_1(k-\kappa)\rangle$ since they differ by a reciprocal lattice vector. The transition between $|\psi_1(k)\rangle$ and $|\psi_1(k-\kappa)\rangle$ then involves a shift in momentum, which is associated with absorption or emission of a quantum of sound mode ($\Omega,\kappa$).

	\begin{figure*}[t]
		\centering
		\includegraphics[width=18cm]{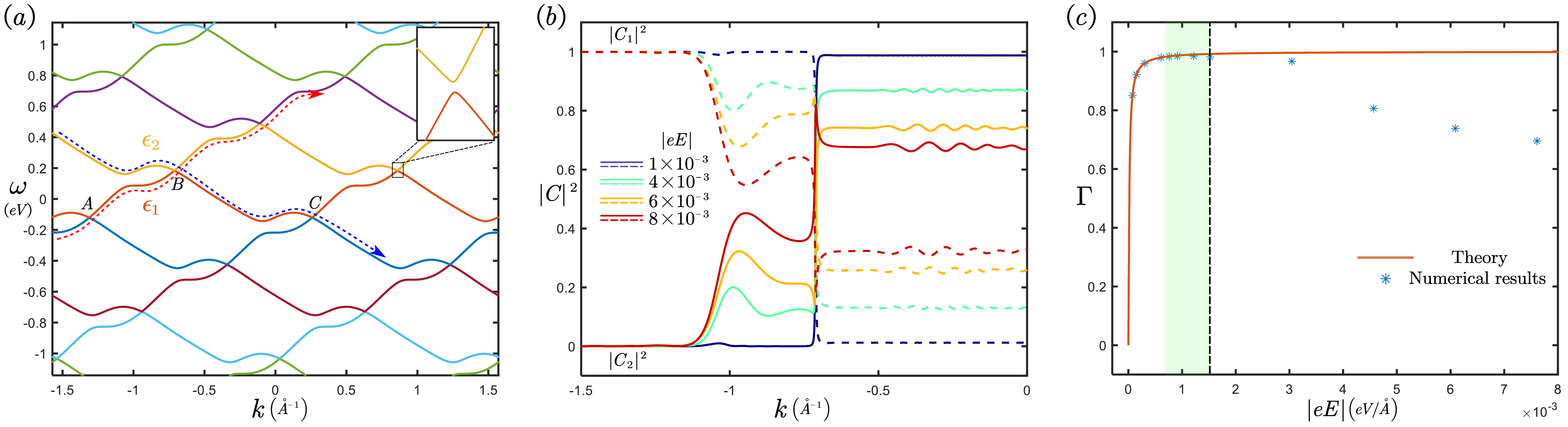}
		\caption{(a) Floquet-Bloch band structure with a very small gap of avoided crossing, calculated for the example of an oscillating Dirac comb with oscillation amplitude $A=0.15$~\cite{SM}, with $A$, $B$ and $C$ labeling three typical points. The blue and red dotted curves represent two paths with topologies also illustrated in Fig.$\ref{torus}$(d). (b) The numerical results of the tunneling process between two bands near the point $B$ in panel (a) by solving Eq.\eqref{evolution} with initial conditions $C_1=1$ and $C_2=0$ under different external field strengths. (c) The numerical results (stars) and the theory from Eq.\eqref{tunneling_rate} (curve) of the tunneling rate of the electron from band 1 to 2. The green shade highlights the area where the tunneling rate is approximately one.} \label{numerical_result_tunneling}
	\end{figure*}

	To make the tunneling happen, we apply an electric field $E$, so that, from the equations of motion in Eq.\eqref{EOM}, electrons will move adiabatically along $k$ axis. The real wavefunction can be approximated as a linear combination of two eigenstates: $|\Phi(t)\rangle = C_1(t)|\psi_1(k(t))\rangle + C_2(t)|\psi_2(k(t))\rangle$. In the context of oblique spacetime crystals, such expansion is valid under a vertical (or irrational) basis representation~\cite{SM}.
	
	Now, plugging the wavefunction $|\Phi(t)\rangle$ into the time-dependent Schrodinger Equation, we obtain the following differential equation regarding $C_{1,2}$:
	\begin{equation}\label{evolution}
		i\hbar\frac{\partial}{\partial t}\left[ \begin{matrix}
			C_1\\
			C_2\\
		\end{matrix} \right] -eE\left[ \begin{matrix}
			\mathcal{A}_{11}&		\mathcal{A}_{12}\\
			\mathcal{A}_{21}&		\mathcal{A}_{22}\\
		\end{matrix} \right] \left[ \begin{matrix}
			C_1\\
			C_2\\
		\end{matrix} \right] =\left[ \begin{matrix}
			\epsilon_1&		0\\
			0&		\epsilon_2\\
		\end{matrix} \right] \left[ \begin{matrix}
			C_1\\
			C_2\\
		\end{matrix} \right]
	\end{equation}
	where $\mathcal{A}_{ij}\equiv \langle\langle\psi_i(k(t))|\left(i\partial_k+x\right)|\psi_j(k(t))\rangle\rangle$ is the multiband Floquet-type Berry connection. This result has the same form as in Bloch crystals but with modified Berry connections. For spacetime crystal, $\mathcal{A}_{ij}$ generically has two contributions:
	\begin{equation}
		\mathcal{A}_{ij} = i\sum_n (f_j)^* \partial_{k} f_i + \sum_n (f_j)^* f_i\mathcal{A}_{k+n\kappa}
	\end{equation}
	where $f_1\rightarrow f_{k}^n$ and $f_2\rightarrow f_{k-\kappa}^{n+1}$.
	The first term is the Floquet contribution, while the second term is the modified Bloch contribution with $\mathcal{A}_{k+n\kappa}$ being the usual Berry connection. For the Floquet-Bloch system generated by a single Bloch band well-separated from all other bands, this Bloch contribution is numerically small and negligible. Then $\mathcal{A}_{ij}(k)$ has only the Floquet contribution that comes solely from the time variations, which allows us to consider only the kernel $\mathcal{H}(k)$ in Eq.\eqref{kernel}.
	
	We again use the oscillating Dirac comb but now with a small gap as an example to show some numerical results. Fig.\ref{numerical_result_tunneling}(a) shows the band structure of such system, which resembles a so-called Landau-Zener grid~\cite{gagge2018bloch,demkov1995crossing}. We then numerically solve Eq.\eqref{evolution} near the gap at point $B$ in Fig.\ref{numerical_result_tunneling}(a), with the electron initially sitting on band $\epsilon_1$ ($C_1=1,C_2=0$). The squared moduli $|C_{1}|^2$ and $|C_{2}|^2$ as functions of $k$ under different external field strengths are plotted in Fig.\ref{numerical_result_tunneling}(b) using dashed and solid curves, respectively. We can see that when $|eE|=10^{-3}eV/\AA$ (blue curves), the evolutions of $|C_{1,2}|^2$ are close to step functions indicating total tunneling through the gap, while for larger $|eE|$, the electron is in a superposition of two bands, violating the adiabaticity. Such violation comes from a larger direct gap near the gap at point $B$, which mixes two bands too early. This tells us that when $|eE|$ is small enough, we can just ignore the influences of that larger gap and only consider the behavior of electron at the vicinity of point $B$, allowing us to have an analytic discussion.

    The system near the point $B$ can be asymptotically approximated by a 2-level system as
    \begin{equation}
			h(k) =  \begin{bmatrix}
				\mathcal{E}_2(k) & \Delta_0/2  \\
				\Delta_0/2 & \mathcal{E}_1(k)
			\end{bmatrix},
	\end{equation} 
	where $\Delta_0$ is the gap at $k=k_B$, and $\mathcal{E}_{1,2}(k)=\mu_{1,2}(k-k_B)$ are the asymptotes of bands $\epsilon_1$ and $\epsilon_2$ near the gap. Thus, we end up with Zener's original tunneling model with a transition rate~\cite{zener1932non}:
	\begin{equation}\label{tunneling_rate}
		\Gamma = \exp\left( -\frac{\pi\Delta_0^2}{2eE|\mu_1-\mu_2|} \right).
	\end{equation}	
	In Fig.\ref{numerical_result_tunneling}(c), we compare the numerical results with the Eq.\eqref{tunneling_rate}, which are in good agreement with each other when $|eE|\leq 1.5\times10^{-3}eV/\AA$. However, as $|eE|$ getting bigger, the discrepancies occur due to the non-adiabaticity of the states before reaching the gap at point $B$.
	
	As discussed in last section, for band structure in Fig.\ref{numerical_result_tunneling}(a) (or Fig.\ref{torus}(d) if $\Delta_0=0$) to have separate topologies, we need the adiabatic condition when electrons are away from the gap (or the band crossing point) and total tunneling when passing through the gap (or the crossing point), which requires the tunneling rate to be one at the point $B$ and zero elsewhere. That can be realized when $|eE|$ is in the green shaded area in Fig.\ref{numerical_result_tunneling}(c).

	\textit{\textbf{Discussion}.---} 
    The uniqueness of the gapless band structure in Fig.\ref{torus}(d) or the similar one but with a tiny gap in Fig.\ref{numerical_result_tunneling}(a) turn out to have non-trivial electric properties that a non-zero DC current can be induced by an external electric field, and the energy can be transferred between the electric field and the sound wave in quantized unit $\Omega$. Thus we can design a prototypical quantum acousto-electric generator based on such system.
	
	Given the band structure depicted in Fig.$\ref{numerical_result_tunneling}$(a), we apply an electric field with strength $|eE|$ lying precisely within the area where $\Gamma\sim 1$. The electrons can then move freely along the red or the blue dotted curves in Fig.$\ref{numerical_result_tunneling}$(a) depending on their initial positions. Those two paths correspond to the band dispersion depicted in Fig.\ref{torus}(b,c), which means that the electrons are oscillating with periods of $\frac{\hbar}{eE}\kappa$ and $\frac{\hbar}{eE}(G-\kappa)$. However, unlike the ordinary Bloch oscillation, the electrons moving along the red or the blue paths actually have nonzero displacements in real space due to the energy change after each period. To see that, we have
	\begin{equation}
		\Delta x = \int \dot{x}dt = \int\frac{\partial \omega(k)}{\partial k}\frac{dk}{\dot{k}} = -\frac{\hbar}{eE}\Delta\omega
	\end{equation}
	where we have applied the equations of motion Eq.$\eqref{EOM}$. The displacements then give a gain of electric energy $\Delta\mathcal{E}=-(-e)E\Delta x=-\hbar\Delta\omega$.
	
	To see how the energy is transferred between the electric field and the sound wave in a quantized unit $\Omega$, we now restrict our consideration within the Brillouin zone which is the energy dispersion from point $A$ to $B$ and to $C$ shown in Fig.$\ref{numerical_result_tunneling}$(a). We have to keep in mind that this system only has one band and all others are just replicas. Imaging one electron sitting initially on the segment $BC$ and driven adiabatically from point $B$ to $C$ by $|eE|$, the energy gain from the electric field is $-\hbar(\omega_C-\omega_B)$. Then at point $C$, the electron will tunnel through the gap to an adjacent state on the lower band which is equivalent to the state at point $B$, since they differ by a reciprocal lattice vector. In other words, this is tunneling from point $C$ to point $B$ on a single band associated with absorption of a quantum of sound mode (changes in both energy and momentum), which is the essence of the intraband Zener tunneling. By oscillating through $B\to C\to B$, the energy is continuously transferred from the sound wave to the electric energy of the electron. When the electron is initially at $AB$, the process is similar but reversed.

	The oscillation periods of those two processes can also be used as experimental signatures to determine whether one of those is happening. Similarly, due to the relation between electronic current and electron velocity: $j=e\rho\dot{x}$ ($\rho$ being the electron density), the two different periods also correspond to different frequencies in the AC part of the current $j$. 
	
	\textit{Acknowledgment.---} The work is supported by NSF (EFMA-1641101) and Welch Foundation (F-1255).

	\clearpage
	
	\begin{widetext}
		\section{Supplementary Material}

\subsection{Oscillating Dirac Comb}
	\begin{figure*}[t]
		\centering
		\includegraphics[width=18cm]{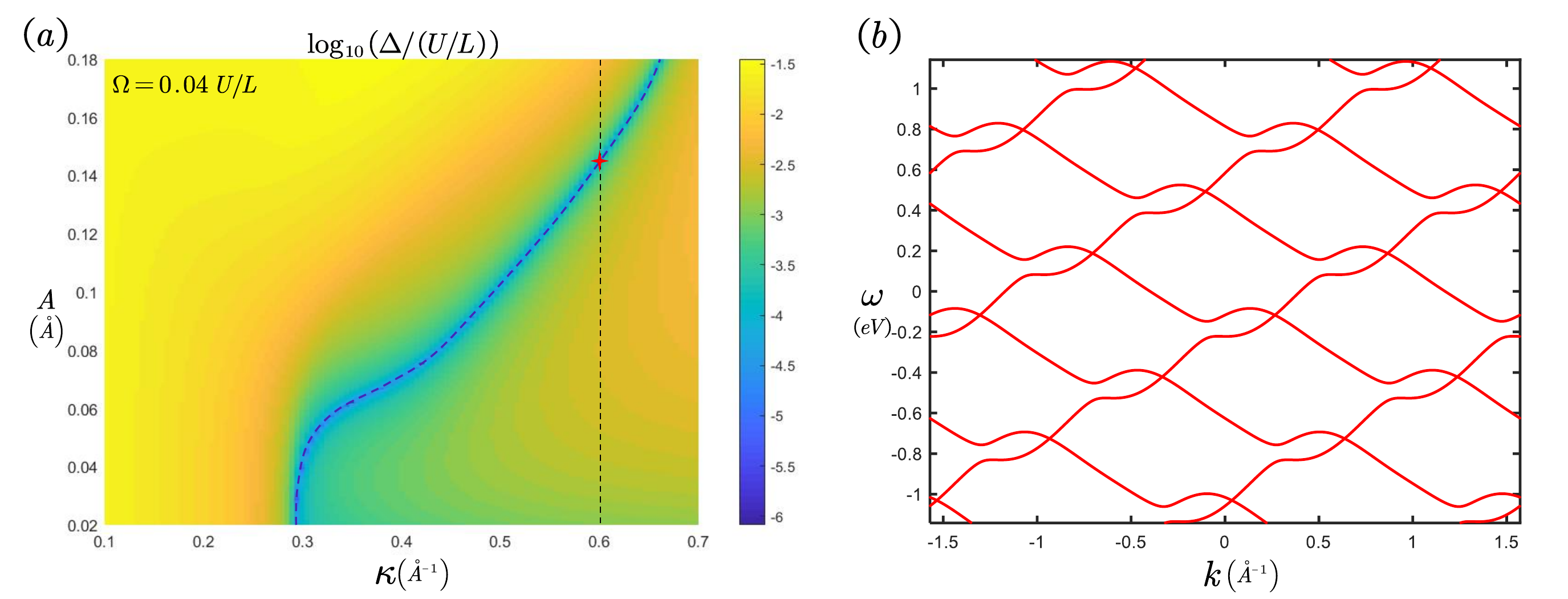}
		\caption{(a) The color map of the logarithm of the bandgap $\Delta$ ($\log_{10} \Delta/(U/L)$) as a function of the oscillating amplitude $A$ and the phonon wavelength $\kappa$, where we have fixed the phonon frequency to be $\Omega=0.04U/L$. The blue dashed line emphasizes the parameters which give gapless band structures. $U/L=$ 7.62eV. (b) The band structure corresponding to the parameters labeled by the red star in panel (a), which is gapless.}\label{gap_close}
	\end{figure*}
	Here we show a toy model called an Oscillating Dirac Comb where the electron is trapped in one-dimensional periodic Dirac potentials that are also shaking periodically in time. The Hamiltonian for such system is:
	\begin{equation}\label{diraccombham}
		\begin{split}
			&H(x,t)=\frac{-\hbar^2}{2m}\partial_{x}^{2}-U\sum_{p} \delta\left(x-pa+A\cos(\kappa pa-\Omega t)\right) 
		\end{split}
	\end{equation}
	where $U$ is a parameter. The corresponding static Hamiltonian at $p$th site is
	\begin{equation}
		H_{0}^{p}=\frac{-\hbar^2}{2m}\partial_{x}^{2}-U\delta(x-pa)
	\end{equation}
	which has only one bound state with energy $E=-U/2L$ and wavefunction $|\phi(x;p)\rangle=(1/\sqrt{L})e^{-|x-pa|/L}$ ($L\equiv\hbar^2/mU$ being typical length). 
    Now we can evaluate the kernel matrix $\mathcal{H}_{m,m}$ up to the second order in $A$ using the tight-binding method:
	\begin{equation}
		\begin{split}
			&\mathcal{H}_{m,m}(k) =-\hbar m\Omega +2\cos((k+m\kappa)a) \beta_{+a},
		\end{split}
	\end{equation}
	\begin{equation}
		\begin{split}
			&\mathcal{H}_{m,m+1}(k) =A\left(\delta^*_{1}+e^{i(k+m\kappa)a} \gamma^*_{-a,1}+e^{-i(k+m\kappa)a} \gamma^*_{+a,1}\right)
		\end{split}
	\end{equation}
	\begin{equation}
		\begin{split}
			\mathcal{H}_{m,m+2}(k) =A^2\left(\sigma^*_{2}+e^{i(k+m\kappa)a} \eta^*_{-a,2}+e^{-i(k+m\kappa)a} \eta^*_{+a,2}\right)
		\end{split}
	\end{equation}
	where we have introduced the spacetime hopping integrals $\beta$, $\delta$, $\gamma$ in first order and $\sigma$, $\eta$ in the second order. They are evaluated as
	\begin{equation}\label{parameters}
		\begin{split}
			\beta_{+a} = &-\frac{U}{L}e^{-\frac{a}{L}}; \\
			\delta_{1} \approx & -i\frac{2U}{L^2}\sin (\kappa a)e^{-2a/L}; \\
			\gamma_{+a,1} \approx & -\frac{U}{2L^2}e^{-\frac{a}{L}}(e^{i\kappa a}-1)=e^{i\kappa a}\gamma_{-a,1}; \\
			\sigma_{2} = & g(L,b)\frac{U}{L^3} - \frac{U}{L^3}\cos(2\kappa a)e^{-\frac{2a}{L}}; \\
			\eta_{+a,2}= &\left( g(L,b)\frac{U}{2L^3} - \frac{U}{8L^3} \right)e^{-\frac{a}{L}}(1+e^{i2\kappa a})=e^{i2\kappa a}\eta_{-a,2},
		\end{split}
	\end{equation}
	where $g(L,b)$ is a parameter that corrects the divergence induced by the unrealistic $\delta$-shape potential.
	The reason why we need to discuss the second order in $A$ is that in this specific model, the leading order is actually the second order.

	In Fig.$\ref{gap_close}$(a), we show how the bandgap of the oscillating Dirac comb depends on the oscillating amplitude $A$ and the phonon wavelength $\kappa$, where we have fixed other parameters: $a=4\AA$, $L =1\AA$, $U=7.62\text{eV}\cdot\AA$, $\hbar\Omega=0.04U/L$, $b = 0.3$, $g = 1.2$. As we can see, the bandgap is most of the time none-zero, but there is a curve (depicted using a blue dashed line) in such parameter space along which the gap is zero. In Fig.$\ref{gap_close}$(b), a typical gapless band structure is depicted with the parameters labeled by a red star in Fig.$\ref{gap_close}$(a), which is $A=0.146\AA$ and $\kappa=0.6\AA^{-1}$. We note that this gap-closing feature is unique in the oblique spacetime crystals where $\kappa\ne 0$, given that the blue dashed curve never intersects with the line $\kappa = 0$ which corresponds to rectangular spacetime crystals. In the main text, we also consider a slightly gaped system in the discussion about the intraband Zener tunneling, where we have changed the oscillating amplitude from $A=0.146\AA$ to $A=0.15\AA$.

		\subsection{Vertical Basis and Irrational Sampling in $k$-space}
		In the (1+1)D system, after solving the Hamiltonian by diagonalizing the kernel matrix $\mathcal{H}_{m,n}$, we find all the eigenenergies and correspondingly the eigenstates. Then we can use these eigenstates as our new basis (perturbed basis, as opposed to the unperturbed Floquet-Bloch basis discussed in the main text). Due to the unique structure of the oblique spacetime crystal, we can have two different choices for representing the basis: the normal basis and the vertical basis.

		As shown in Fig.~\ref{vertical_basis}, two different $K$-space sampling methods are depicted. In the left panel, we use the most usual way of sampling which is to choose a set of equally spacing $K$-points within the first Brillouin zone (in this case, the $n=0$ band), and then an arbitrary state can be expanded as
		\begin{equation}\label{nbasis}
		    |\Psi\rangle = \sum_{k\in\mathcal{N}} C_k(t)|\Phi_{\omega_0(k)}(x,t)\rangle,
		\end{equation}
		where $\mathcal{N}$ is the set of equally spacing $K$-points, and $C_k(t)$ is the time-dependent coefficient. However, if the system has an irreducible oblique structure where $\kappa/G$ is an irrational number, we can have another way of sampling the $K$-space, and it is shown in the right panel of Fig.~\ref{vertical_basis}. A vertical line intersects with all bands with different index $n$ giving the crossing points exactly the same momentum $k$ (an arbitrarily selected $k$). But because the periodicity of the oblique structure, we can transform all crossing points back into the first Brillouin zone by doing $(k+n\kappa)\mod G$ for all $n$ (see, for example, the black arrows in the right panel of Fig.~\ref{vertical_basis}). Because of the irrationality of $\kappa/G$, the process of $(k+n\kappa)\mod G$ can densely and almost evenly sample the $K$-space within the first Brillouin zone when $n$ is large. That is called irrational sampling. As a result, the state $|\Psi\rangle$ can also be expanded as
		\begin{equation}\label{vbasis}
		    |\Psi\rangle = \sum_{n} C_n(t)|\Phi_{\omega_n(k)}(x,t)\rangle \equiv \sum_{k'\in\mathcal{V}_k} C'_{k'}(t)|\Phi_{\omega_0(k')}(x,t)\rangle,
		\end{equation}
		where $\mathcal{V}_k = \{ k'|k'= (k+n\kappa)\mod G \text{, for } \text{all } n\}$ is the set of $K$-points in the first Brillouin zone generated by irrational sampling. We have to note that $\mathcal{V}_k\ne\mathcal{N}$, but when $n$ is large, those two are equivalent to each other.
		
		For some practical purposes, it is more convenient to use the vertical basis since it favors the conservation of the quasimomentum. In the main text, we used this vertical basis and only considered two coefficients that contribute the most to the intraband Zener tunneling.
		\begin{figure*}[t]
			\centering
			\includegraphics[width=17.5cm]{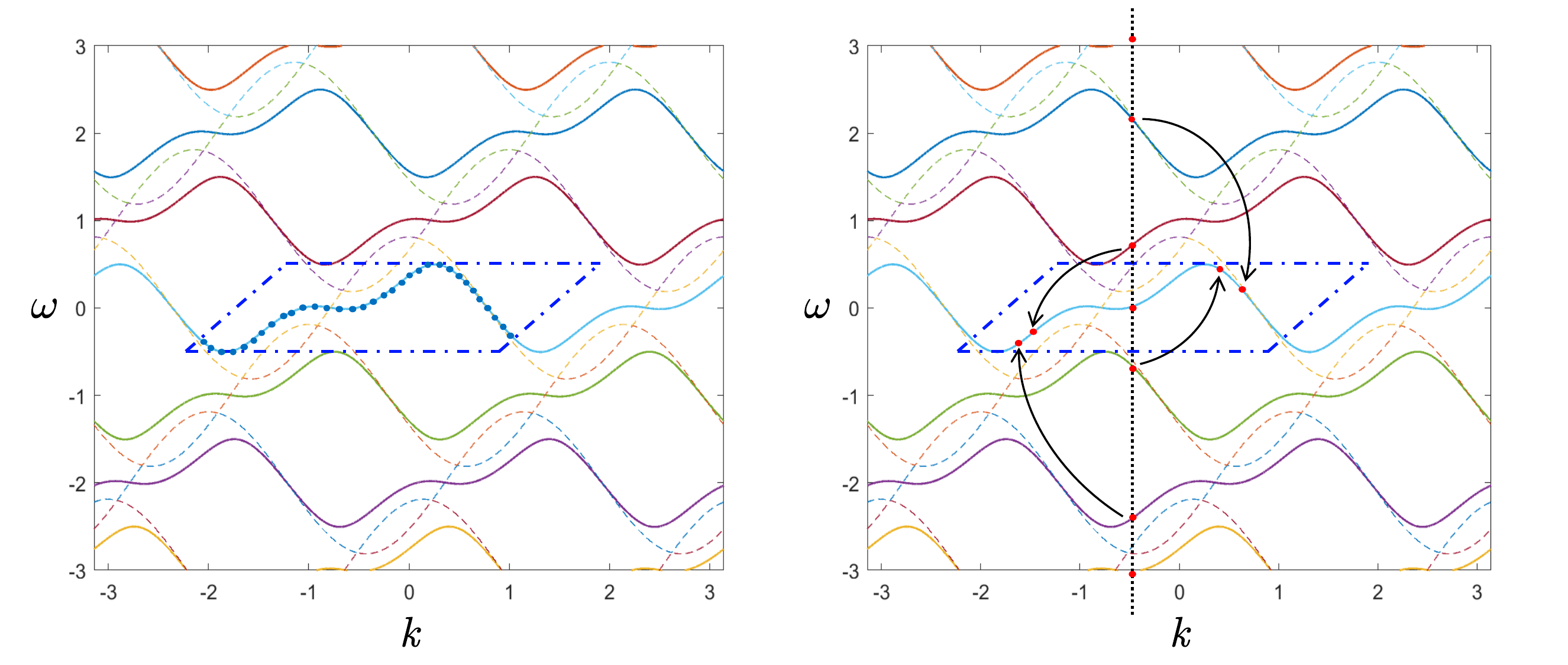}
			\caption{Two different bases: (left panel) the normal basis with the blue dots denoting the equally spacing $K$-point sampling within one Brillouin zone; (right panel) the vertical basis with the red dots denoting the vertical (irrational) $K$-point sampling along a vertical line.}\label{vertical_basis}
		\end{figure*}
		
		\subsection{Intraband Zener Tunneling}\label{AppendixDerivation}
		Fig.\ref{tunneling} shows a schematic plot of the intraband Zener tunneling process, where two states $|\psi_1(k(t))\rangle $ and $|\psi_2(k(t))\rangle $ setting on two adjacent replicas.
		After applying the electric field, the Hamiltonian becomes $\tilde{H} = H + eEx$ where $H$ is the original Floquet-Bloch Hamiltonian. By plugging the wave-function into the Schrodinger Equation $\tilde{H}|\Phi(t)\rangle = i\hbar\partial_t |\Phi(t)\rangle$, we get
		\begin{equation}
			H|\Phi(t)\rangle = (i\hbar\partial_t-eEx)|\Phi(t)\rangle
		\end{equation}
		Then, by utilizing the equation of motion $\dot{k} = -eE/\hbar$, we have
		\begin{equation}
			\begin{split}
				&C_1(t)\omega(k(t))|\psi_1(k(t))\rangle + C_2(t)(\omega(k(t)-\kappa)+\Omega)|\psi_2(k(t))\rangle \\
				=& i\hbar\frac{\partial C_1}{\partial t}|\psi_1(k(t))\rangle + C_1(t)eE\left(-i\frac{\partial}{\partial k}-x\right)|\psi_1(k(t))\rangle \\
				+& i\hbar\frac{\partial C_2}{\partial t}|\psi_2(k(t))\rangle + C_2(t)eE\left(-i\frac{\partial}{\partial k}-x\right)|\psi_2(k(t))\rangle.
			\end{split}
		\end{equation}
		We have the orthonormal relation between two eigenstates: $\langle\langle\psi_i(k)|\psi_j(k)\rangle\rangle = \delta_{ij}$, but to avoid the integration over time, we can see that the orthonormal relation also holds in space that $\langle\psi_i(k)|\psi_j(k)\rangle=\delta_{ij}$ due to the momentum conservation. Then we can obtain two differential equations for $C_{1,2}$:
		\begin{equation}\label{evolution_o}
			\begin{split}
				C_1(t)\omega(k(t)) =& i\hbar\frac{\partial C_1}{\partial t} + C_1(t)eE\langle\psi_1(k(t))|\left(-i\frac{\partial}{\partial k}-x\right)|\psi_1(k(t))\rangle \\ 
				&+ C_2(t)eE\langle\psi_1(k(t))|\left(-i\frac{\partial}{\partial k}-x\right)|\psi_2(k(t))\rangle\\
				C_2(t)(\omega(k(t)-\kappa)+\Omega) =& i\hbar\frac{\partial C_2}{\partial t} + C_1(t)eE\langle\psi_2(k(t))|\left(-i\frac{\partial}{\partial k}-x\right)|\psi_1(k(t))\rangle\\
				&+ C_2(t)eE\langle\psi_2(k(t))|\left(-i\frac{\partial}{\partial k}-x\right)|\psi_2(k(t))\rangle,
			\end{split}
		\end{equation}
		where one can also show that due to the conservation of momentum, the factors in above equation actually satisfy
		\begin{equation}
		    \langle\psi_i(k(t))|\left(-i\frac{\partial}{\partial k}-x\right)|\psi_j(k(t))\rangle = \langle\langle\psi_i(k(t))|\left(-i\frac{\partial}{\partial k}-x\right)|\psi_j(k(t))\rangle\rangle.
		\end{equation}
		Thus, we can recast the Eq.\eqref{evolution_o} into a Matrix form:
		\begin{equation}
			i\hbar\frac{\partial}{\partial t}\left[ \begin{matrix}
				C_1\\
				C_2\\
			\end{matrix} \right] -eE\left[ \begin{matrix}
				\mathcal{A}_{11}&		\mathcal{A}_{12}\\
				\mathcal{A}_{21}&		\mathcal{A}_{22}\\
			\end{matrix} \right] \left[ \begin{matrix}
				C_1\\
				C_2\\
			\end{matrix} \right] =\left[ \begin{matrix}
				\epsilon_1&		0\\
				0&		\epsilon_2\\
			\end{matrix} \right] \left[ \begin{matrix}
				C_1\\
				C_2\\
			\end{matrix} \right]
		\end{equation}
		where $\mathcal{A}_{ij}\equiv \langle\langle\psi_i(k(t))|\left(i\partial_k+x\right)|\psi_j(k(t))\rangle\rangle$ is the multiband Floquet-type Berry connection.
		
	\begin{figure}[t]
		\centering
		\includegraphics[width=7.75cm]{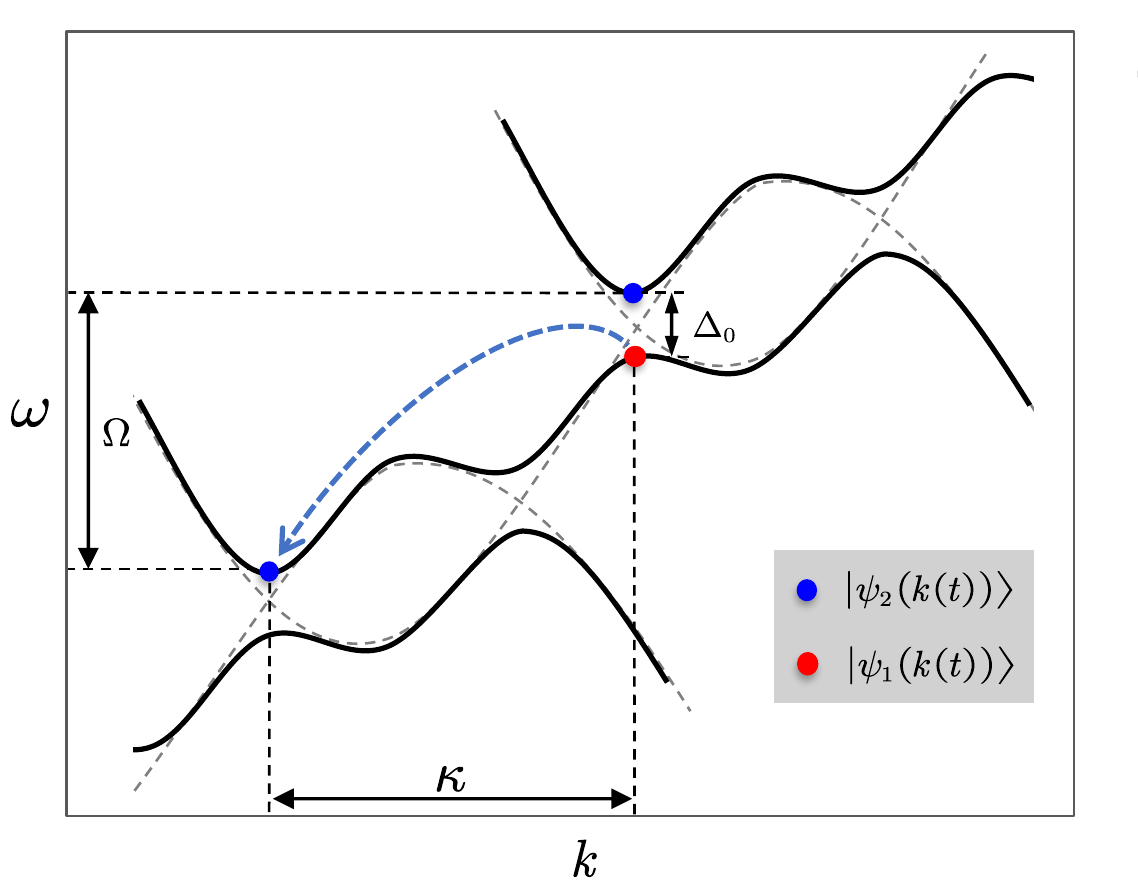}
		\caption{Zener tunneling across a small gap $\Delta_0$ occurs when the momentum $k$ changes with time under an external field just like the tunneling in Bloch states.  However, since the different bands are really replica of the same Floquet-Bloch band, shifted by the phonon energy and momentum $(\Omega, \kappa)$, Zener tunneling in the present context is actually an intraband process indicated by the dashed arrow. }\label{tunneling}
	\end{figure}
	
	\end{widetext}

\begin{thebibliography}{35}
\bibitem{footnote1}
		Due to the convention of quantum mechanics that we always write the propagating phase factor as $e^{i\boldsymbol{k}\cdot\boldsymbol{x}-i\omega t}$, the space and time have naturally different signatures encoded. In this paper, the signature is set to be $(+,+,+;-)$ where the last entry stands for time. Then we can check that the reciprocal lattice vectors indeed fulfill: $-(\boldsymbol{0},\frac{2\pi}{\Omega}) \cdot(\boldsymbol{\kappa},\Omega)=(\boldsymbol{R},\frac{\boldsymbol{\kappa}\cdot\boldsymbol{R}}{\Omega})\cdot(\boldsymbol{G},0)=2\pi$ and $(\boldsymbol{0},\frac{2\pi}{\Omega}) \cdot(\boldsymbol{G},0)=(\boldsymbol{R},\frac{\boldsymbol{\kappa}\cdot\boldsymbol{R}}{\Omega})\cdot(\boldsymbol{\kappa},\Omega)=0$
\bibitem{footnote2}
		The system for obtaining the typical Floquet-Bloch band structure in Fig.\ref{band} is a kernel Matrix in Eq.\eqref{kernel} with the diagonal elements being shifted `cos'-shape band dispersion: $\mathcal{H}_{n,n}=2\beta\cos [(k+n\kappa)a]-n\Omega$, and with the only nonzero off-diagonal terms being $\mathcal{H}_{n,n\pm1}=\delta$. The parameters used are $a=2$, $\kappa=\Omega=1$, $\beta=0.4$, and $\delta=0.3$.
\bibitem{footnote3}
		This is equivalent to say that $\omega_n(k)=\omega_n(k+\kappa)-\Omega=\omega_{n+1}(k)$ which means that $\omega_n(k) \equiv\omega_0(k)$ for all $n$. Later, we will discuss that it requires a very unique band topology depicted in Fig.\ref{torus}(b).
\bibitem{footnote4}
		Fig.\ref{torus}(b) represents a band topology with $\omega_n(k+\kappa)=\omega_n(k)+\Omega$. If $\kappa=0$ which corresponds to a rectangular spacetime crystal, then there must be a point on the band such that $\dot{x}=\frac{\partial \omega_n(k)}{\partial k}\to\infty$, which is forbidden by mechanics.
\end{thebibliography}
\end{document}